\documentclass[aps,prl,twocolumn,superscriptaddress]{revtex4}
\usepackage{epsfig}
\usepackage{graphicx}
\usepackage{bm}

\newcommand{\be}{\begin{equation}}

\newcommand{\ee}{\end{equation}}
\newcommand{\bea}{\begin{eqnarray}}
\newcommand{\eea}{\end{eqnarray}}

\begin{document}
\title{A Classification Scheme for Phenomenological Universalities in Growth Problems }
\author{P.Castorina$^{(a,b)}$, P.P.Delsanto$^{(c,d,e)}$, C.Guiot$^{(d,f)}$}
\affiliation{$^{(a)}$ Department of Physics, University of
Catania,
Italy\\  $^{(b)}$ INFN-Catania, Italy  \\
$^{(c)} $ Department of Physics, Politecnico di Torino,  Italy \\ $^{(d)} $ CNISM, Sezioni di Torino Universita' e Politecnico, Italy \\ 
$^{(e)} $ Bioindustry Park of Canavese, Ivrea, Italy\\
$^{(f)}$ Department of Neuroscience, Universita' di Torino, Italy}

\begin{abstract}

A classification in universality classes of broad categories of phenomenologies, belonging to different disciplines,
 may be very useful for a crossfertilization among them and for the purpose of pattern recognition. 
We present here a simple scheme for the classification of nonlinear growth problems. 
The success of the scheme in predicting and characterizing the well 
known Gompertz, West and logistic models, suggests to us the study of a hitherto unexplored class of nonlinear growth problems.

\end{abstract}

\maketitle

Present efforts towards the understanding of complex systems in physics, biology, economics and social science require complementary microscopic and macroscopic
descriptions. In fact, due to the complexity of the underlying dynamics and the unbounded  variety of external conditions, a fundamental approach is missing.
 Microscopic models depend on such a large number of parameters that they often lose almost any predictive power, even when the calculations do not become forbiddenly 
difficult or time consuming. On the other hand, macroscopic descriptions are often inadequate and do not take advantage of the enormous progress that has been achieved
 at the microscopic level in recent years.
 An intermediate (mesoscopic) approach \cite{pp1,pp2,pp3} may be very fruitful, but a bridging among the various levels \cite{pp4} is not always
 easy to accomplish.

A different approach has consequently emerged for the treatment of problems
, which do not directly require a detailed description of the system to be investigated. 
The idea is to exploit the spectacular advancement of interdisciplinary research, which has 
taken place in the last two decades or so, involving e.g. the relevance of scale laws, complexity and nonlinearity in virtually all disciplines.
	
In this context many patterns have been discovered, which are remarkably similar, although they concern completely different phenomenologies. 
This is hardly surprising, since often the "background" mathematics is the same. We shall call them ``phenomenological universalities" \cite{pp5}, 
in the sense that they refer to a "transversal" generality (not to a uniformly general 
behaviour within a given class of phenomena). 

As examples of universality we can quote 
the ``life's universal scaling laws'' \cite{westa}, which will be discussed later, 
and the ``universality of nonclassical nonlinearity'' \cite{pp6}. The latter suggests 
that unexpected effects, such as those recently discovered by P. Johnson and collaborators \cite{john} and 
called by them ``Fast Dynamics'', may be found as well, although possibly with quite different
 manifestations in other fields of research.

A reliable macroscopic analysis of a complex system requires two fundamental ingredients: non linearity and stochasticity.
 Non linearity is  more fundamental because the stochastic behaviour requires a  non linear dynamics.
 Therefore non linearity must be considered as the fundamental feature of these systems and in this letter we consider general growth 
problems based on this crucial aspect. We shall show that different ``degrees of nonlinearity'' (as specified below) 
correspond to various growth patterns, which can be systematically classified.

For this purpose, let us consider the very broad class of growth phenomena, which may be described by the simple law:
\be
\frac{dY(t)}{dt} = \alpha(t) Y(t)
\ee
where $\alpha(t)$ represents the specific growth rate, which may vary with time, of a given variable $Y(t)$. By introducing the
nondimensional variables $\tau=\alpha(0) t$, $y(t)=Y(t)/Y(0)$ and $a(\tau)=\alpha(t)/\alpha(0)$,
Eq.(1) becomes :
\be
\frac{dy(\tau)}{d\tau} = a(\tau) y(\tau)
\ee
with $ y(0) = a(0) = 1$.
By defining the time variation of $a(\tau)$ through a function $\Phi(a)$:
\be
\Phi(a)= - \frac{da(\tau)}{d\tau}
\ee
we obtain a system of two differential equations, which may generate a variety of growth patterns, according to the explicit form of $\Phi(a)$ ,
 and is usually analyzed by the standard fixed points and characteristic curves methods \cite{tri}.

In this contribution we are not directly interested in this aspect, but we wish to show, instead, how the nonlinear terms in $\Phi(a)$
 affect the growth dynamics process.

We assume that $| a(\tau) | < 1$ and expand $\Phi(a)$ in power series
\be
\Phi(a)= \Sigma_{n=0}^\infty b_n a^n
\ee
in which we retain only a limited number  of $N+1$ terms. 
Borrowing from the language of phase transitions \cite{phase} , we define, as belonging to the phenomenological universality class of order N
 (which we shall call UN, N=1,2,...), the ensemble of all the phenomenology, which may be suitably described by truncating the series at 
the power n=N. In the following we shall analyse in detail the classes U1, U2 and U3 and provide a description of their nonlinear properties.

The ``linear''  behaviour of the system corresponds to a constant specific growth rate, i.e. $\Phi(a)=0$ ( or $b_n= 0$ for any $n$).  
Then $y(\tau)$ follows a purely exponential law. Also the case $b_0 \neq 0$  with all $b_n =0$ for $ n\ge 1$, 
can be easily shown to lead to an exponential growth. 
Since we are interested only in the nonlinear effects, we shall assume $b_0= 0$.
 This does not cause any loss of generality, since one can always expand $\Phi$ in the variable $\beta= a -c$, where c is a solution of $\Sigma_{n=0}^\infty b_n c^n=0$.
In the $\beta$ expansion the coefficient of $\beta^0$ vanishes. Likewise, again without any loss of generality , we can set $b_1=1$, 
as one would have from an expansion in the variable $\gamma = a/b_1$.

In order to study the various classes of universality and obtain the corresponding differential equations and solutions, we write from Eqs. (2) and (3):
\be
- \Phi(a) \frac{dy}{da}=ay
\ee

from which it follows:
\be
ln y = - \int \frac{a da}{\Phi(a)} + const
\ee
By solving the previous equation with respect to the variable $a(\tau)$ and then substituting into Eq. (2), one obtain the differential equation characterizing the class.
The integration constant can be easily obtained from the initial conditions.

Let us then start by considering the class U1, i.e. with N=1. From Eq. (6)  and $\Phi(a)=a$ , it immediately follows:
\be
\frac{dy}{d\tau} = y - y ln y
\ee
with the solution
\be
y=exp[1-\exp{(-\tau)}]
\ee
Eq.(7) represents the ``canonical'' form of U1 differential equations and corresponds to the Gompertz law, originally introduced \cite{gomp} in actuarial mathematics 
to evalute the mortality tables and, nowdays, largely applied to describe economical and biological growth phenomena. For example, the Gompertz law gives
a very good phenomenological description of the tumor growth pattern \cite{nort}, \cite{wel} and  it can be related to  the energetic cellular balance \cite{paolo}.
 It is remarkable that it does not contain any free parameter ( except for the scale and linear parameters which have not been included, as discussed before),
 i.e. all Gompertz curves are (under the mentioned proviso) identical.

Let us now turn our attention to the class U2, i.e. N=2. From Eqs. (6) and (3) and 
$\Phi(a)= a + b a^2$, where $b= b_2$ ,  it follows
\be
\frac{dy}{d\tau} =\alpha_2 y^p - \beta_2 y
\ee
where $\alpha_2=(1+b)/b$, $p=1-b$ and $\beta_2=1/b$ with the solution
\be
y=[1+b-b\exp{(-\tau)}]^{1/b}
\ee
By identifying $y$ with the mass of a biological system,$y=m$, and defining the asymptotic mass ($ m_0=y_0=1$)
\be
M = lim_{\tau \rightarrow \infty} m(\tau)= (1+b)^{1/p}
\ee
it is easy to show that Eqs. (9) and (10) correspond to the well known allometric West equation for the case $p=3/4$ \cite{westb}.
 In their ontogenetic growth model, $m$ represents the mass of any living organism, from protozoa to mammalians (including plants as well).
 By redefining their mass and time variables $ z =1- (y/m)^b$
 and $\theta=-\tau + ln b - b ln M$ they obtain the very elegant parameterless universal law
\be
z=exp(-\theta)
\ee
which fits well the data for a variety of different species, ranging from shrimps to hens to cows.
 It is interesting to note that, in a subsequent work \cite{westc}, West and collaborators give an interpretation of $\theta$ as the ``biological time''
, based on the organism's internal temperature.

	An extension of West's law to neoplastic growths has been recently suggested by C. Guiot, P.P. Delsanto, T.S. Deisboeck and collaborators \cite{tom1,tom2}.
 Although an unambigous fitting of experimental data is much harder in tumors (except for cultures '¡Æin vitro'¡Ç of multicellular tumor spheroids),
 the extension seems to work well . Of course, particularly '¡Æin vivo'¡Ç, other mechanisms must be taken into account, such as the pressure from 
the surrounding tissue \cite{cg1}. Another important issue is the actual value of the exponent p, which has been the object of a strong 
debate \cite{debate}. Recently C. Guiot et al. \cite{cg2} have proposed that p may vary dinamically with the fractal nature of the input channels
 (e.g. at the onset of angiogenesis).
	
Although it is not obvious from a comparison between Eq. (7) and Eq. (9), U1 represents a special case $(b=0)$ of U2, as it obviously follow from 
the power expansion of   $\Phi$ (which has $b=0$ in U1). This can  be verified directly  by carefully performing the limit $b \rightarrow 0$ in Eq. (10) .
 In fact it is interesting to plot $y$ vs. $\tau$ in a sort of phase diagram ( see Fig. 1) .

\begin{figure}
\epsfxsize=8truecm \centerline{\epsffile{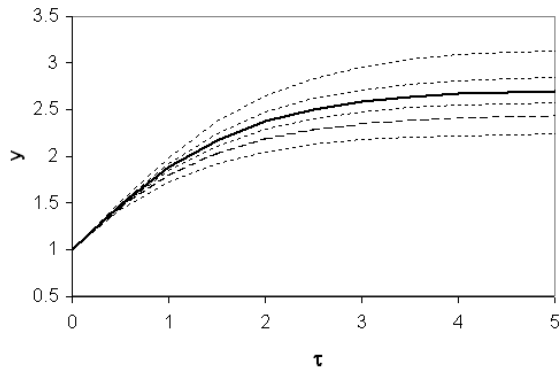}}
\noindent {\caption{\label{fig:1} - {Growth curves belonging to the class U2. From the top to the bottom the values of the
parameter $b$ are $-0.25,-0.1,0.1,0.25,0.5$ respectively. The solid curve ($b=0,p=1$) corresponds to the Gompertzian (U1),
while the dashed one refers to the value proposed in \cite{westa} $p=3/4$ ($b=1/4$).
}}}
\end{figure}

 This leads to a very suggestive interpretation of Eq. (9).
 Having added a term to the  $\Phi(a)$ expansion, we gain, in U2, the possibility of adding a ``new'' ingredient, which turns out to 
be a different dimensionality of the ``energy flux'' i.e. input, output and consumption (metabolism).
 E.g. the first term on the RHS of Eq. (9) may be related \cite{westd}
 to the premise that the tendency of natural selection to optimise energy transport has led to the evolution of fractal-like distribution 
networks with an exponent p for their terminal units vs. an exponent 1 for flux mechanisms related to the 
total number of cells. When b=0, p=1 and we lose the new ingredient, thus falling back into U1.

  This is confirmed also by considering the logistic equations, corresponding to eq.(9) with negative $b$. The usual logistic equation 
is obtained for $p=2$. As well known in population dynamics \cite{roy}, in this case the new ingredient is the  competition for resources.

Finally we consider the class U3. Writing
\be
\Phi(a)= a ( 1+b a + c a^2)
\ee
from Eq. (6)  it follows
\be
\int \frac{ da}{1+ba+ca^2}= K - lny
\ee
In this case there are three subclasses, U31, U32 and U33, corresponding to $\Delta = 4c -b^2 \ge < 0$ .
For brevity we limit ourselves to report here the canonical equation for U31, i.e. when $\Delta<0$:
\be
\frac{dy}{d\tau} =\alpha_3 y - \beta_3  y^p + \gamma_3 \frac{dy^p}{d\tau} 
\ee
where $d= \sqrt{-\Delta}$ ,$ p=1-d$,$ K=(d-3c)/(d+3c)$, $\alpha_3= (d-c)/2c$ and $\beta_3 = K(d+c)/2c$ and $\gamma_3= K/(1-d)$.
It is interesting to observe that, in the same way that U2 adds ( with respect to U1) a term with a different dimensionality to the
 energy flux contribution, U3 adds such a term (the last one in Eq. 15) to the growth part.

	To conclude, we have developed a simple scheme, which allows the classification in nonlinear phenomenological 
universality classes of all the growth problems, which can be described by Eqs. (2) and (3). We have found that the first 
class U1 corresponds to the Gompertz curve, which has no free parameters (apart from scale and linear ones). The second class 
U2 includes all the Westlike and logistic curves and has a free parameter b: when b=0 we fall back into U1 (Gompertz).
	The success of the scheme in obtaining the classes U1 and U2 when one or two terms are retained in the expansion of 
$\Phi(a)$  has suggested to us to investigate the class U3, which is generated by simply adding one more term (see Eq. 13). 
To our knowledge, this class has never been investigated before. A remarkable result is that each new class adds a new ``ingredient'' (or growth mechanism). 
E.g. U2 allows for the possible presence of two dimensionalities in the energy flux. U3 extends such a possibility to the growth term ( the time derivative).
	In addition to its intrinsic elegance \cite{elegance} the concept of universality classes may be useful 
for several reasons of applicative relevance. In fact it greatly facilitates the crossfertilization among different
 fields of research by implicitly suggesting that a method of analysis, which is proven advantageous in one study, be
 tried and eventually adopted in others. Also, if an unexpected effect is found experimentally in a field, similar effects
 ``mutatis mutandis'' should also be sought in similar, although unrelated, experiments in other fields. Finally,
 if a detailed study is performed to recognize the patterns that are characteristics of the most relevant classes (and subclasses)
, this could greatly help in classifying and fitting new sets of experimental data independently of the field of application.

{\bf Acknowledgements}

We wish to thank Drs. M. Griffa and F. Bosia for their help and useful discussions. 
This work has been partly supported by CviT (Centre for the development of a Virtual Tumor).

\end{document}